\documentclass[aps,letterpaper,twocolumn,preprintnumbers,floatfix,superscriptaddress]{revtex4}

\usepackage{graphicx}
\usepackage{dcolumn}
\usepackage{bm}
\usepackage{epstopdf}
\usepackage{epsfig}
\usepackage{gensymb}
\usepackage{mathrsfs}
\usepackage{amsmath}
\usepackage{amssymb}
\usepackage{graphicx,bm}
\usepackage{slashed}
\usepackage[makeroom]{cancel}
\usepackage{dsfont}
 \usepackage[titletoc]{appendix}

\def\fun#1#2{\lower3.6pt\vbox{\baselineskip0pt\lineskip.9pt
  \ialign{$\mathsurround=0pt#1\hfil##\hfil$\crcr#2\crcr\sim\crcr}}}

\def\lsim{\mathrel{\rlap{\raise 2.5pt \hbox{$<$}}\lower 2.5pt\hbox{$\sim$}}}
\def\gsim{\mathrel{\rlap{\raise 2.5pt \hbox{$>$}}\lower 2.5pt\hbox{$\sim$}}}

\input epsf

\newcommand{\be}{\begin{equation}}
\newcommand{\ee}{\end{equation}}
\newcommand{\bea}{\begin{eqnarray}}
\newcommand{\eea}{\end{eqnarray}}

\usepackage{color}

\newcommand{\comment}[1]{}

\begin{document}

\title{Twin Higgs with Exact $Z_2$}

\author{Csaba Cs\'aki}
\affiliation{ Department of Physics, LEPP, Cornell University, Ithaca, NY 14853, USA}

\author{Cong-Sen Guan}
\affiliation{ Department of Physics, LEPP, Cornell University, Ithaca, NY 14853, USA}
\affiliation{
 CAS Key Laboratory of Theoretical Physics, Institute of Theoretical Physics,
Chinese Academy of Sciences, Beijing 100190, China}
\affiliation{School of Physical Sciences, University of Chinese Academy of Sciences, Beijing 100049, P. R. China}
\affiliation{PLA Naval Submarine Academy, Qingdao 266199, P. R. China}

\author{Teng Ma}
\affiliation{
Physics Department, Technion - Israel Institute of Technology, Haifa, 3200003, Israel}
\author{Jing Shu}
\affiliation{
 CAS Key Laboratory of Theoretical Physics, Institute of Theoretical Physics,
Chinese Academy of Sciences, Beijing 100190, China}

\affiliation{School of Physical Sciences, University of Chinese Academy of Sciences, Beijing 100049, P. R. China}
\affiliation{ CAS Center for Excellence in Particle Physics, Beijing 100049, China}

\begin{abstract}
We present a novel mechanism for realistic electroweak symmetry breaking in Twin Higgs/neutral naturalness models where the $Z_2$ exchange symmetry can remain exactly unbroken. The exchange symmetry in the Yukawa sector will be implemented as an "N-trigonometric parity'' $\sin N \frac{h}{f} \leftrightarrow \cos N \frac{h}{f}$. The Yukawa couplings will be suppressed leading to an N-suppressed Higgs quadratic term, without significantly affecting the quartic. We present a concrete implementation of this idea for general (odd) values of N using maximal symmetry, and a realistic benchmark model for $N=3$. We find that the tuning in the resulting Higgs potential is negligible, and also show that two-loop N-suppression violating gauge contributions can be sufficiently small. 
The Higgs potential and its couplings in top sector are different from other neutral naturalness models, which are the main predictions of our model and can be tested in colliders.   
\end{abstract}

\pacs{11.30.Er, 11.30.Fs, 11.30.Hv, 12.60.Fr, 31.30.jp}

\maketitle

\section{Introduction}
Obtaining a natural model of electroweak symmetry breaking (EWSB) is one of the biggest challenges in particle physics. Within the standard model (SM) the Higgs potential is quadratically sensitive to UV scale physics and is not calculable. In supersymmetric or composite Higgs extensions the UV sensitivity is softened to a logarithmic sensitivity, and in some cases the entire Higgs potential can be finite and calculable. This softening is achieved by introducing colored top partners and electroweak gauge partners with equal (composite Higgs) or opposite (supersymmetry) spins. However, direct experimental searches from the LHC are starting to put stringent lower bounds above 1 TeV on these partners, rendering these models tuned at the percent level. A possible way around these bounds is neutral naturalness~\cite{Chacko:2005pe,TH2,Craig,Geller:2014kta,Low:2015nqa,Barbieri:2015lqa}, where the partners responsible for the softening of the Higgs potential are not charged under the SM interactions, and in particular the top partners would not carry color under ordinary QCD, but rather under a hidden ``twin color''. In this case the direct experimental bounds will be much weaker and a natural EWSB can still be hoped for. The most prominent models with neutral naturalness are the Twin Higgs Models (THM)~\cite{Chacko:2005pe,TH2} where a $Z_2$ exchange symmetry enforces the cancellation of the leading UV sensitivities both in the top and the gauge sectors. In these models (as in ordinary composite Higgs models~\cite{Kaplan:1983fs,Georgi:1984af,Dugan:1984hq,ArkaniHamed:2001ca,SILH,Bellazzini:2014yua,Agashe:2004rs,LH}) the Higgs is also a pseudo-Goldstone boson (pNGB) of a spontaneously broken global symmetry. In ordinary THM's the $Z_2$ symmetry has to be broken: without such breaking it is usually very difficult to obtain a hierarchy between the pNGB Higgs VEV $\langle h \rangle $ and the global symmetry breaking scale $f$, $\langle h \rangle/f \ll 1$. Such a $Z_2$ breaking usually reintroduces the quadratic dependence of the Higgs potential on some of the partner masses, and as a consequence brings back some of the tuning (usually of order 10\%).

In this paper we introduce a novel way of generating the $\langle h \rangle/f$ hierarchy in THM's without breaking the $Z_2$ symmetry. The essence will be based on the ``N-suppression mechanism''. We will consider a case where the implementation of the $Z_2$ symmetry in the top sector is in the form
\begin{equation}
\sin \frac{Nh}{f} \leftrightarrow \cos \frac{Nh}{f}
\end{equation}
for some odd integer $N$.  We will see that this will imply that the top Yukawa couplings depend on $\sin  \frac{Nh}{f}$ or $\cos  \frac{Nh}{f}$, strongly reducing the value of the Yukawa coupling. This will result in a $1/N^2$ suppression of the quadratic term in the Higgs potential, while the quartic will be essantially unchanged. As a result the structure of the Higgs potential will be modified, and generating the $\langle h \rangle/f$ hierarchy (and hence natural EWSB) will be achieved without significant tuning. The exact $Z_2$ symmetry in this model will result in light hidden fields, such as the twin photon and twin neutrinos, which generically result in significant contributions  to the radiation density and $N_{eff}$, in conflict with recent cosmological observations~\cite{Ade:2015xua}. This tension can be resolved by breaking the $Z_2$ symmetry
in the light lepton sector, for example via mixings among the neutrinos and twin neutrinos~\cite{ Csaki:2017spo}. This can lower the decoupling temperature between the two sectors while giving negligible $Z_2$ breaking contributions to the Higgs potential. For other ways to obtain realistic cosmologies in mirror THM's see~\cite{Craig:2016lyx}.

The paper is organized as follows. In section~\ref{Sec:N-suppression} we present the essence of the N-suppression mechanism. We analyze the consequences on the Higgs potential and show what limit is needed in order to have a realistic $t\bar{t}h$ coupling for this model. In Sec.~\ref{sec:Fermion_sector} we show how to implement our N-suppression mechanism in a concrete model using additional Dirac fermions and maximal symmetry, and then focus our attention on the $N=3$ realistic benchmark model. The structure of the Higgs potential and the analysis of the tuning is presented in Sec.~\ref{sec:tuning}, while we comment on the magnitude of two-loop N-suppression violating effects in Sec.~\ref{sec:twoloop} and on novel aspects of phenomenology in Sec.~\ref{sec:pheno}. We conclude in Sec.~\ref{sec:conclusions}, while the Appendices we present the gauge sector of the model (App.~\ref{App:gauge}) as well as the discrete symmetries needed to ensure N-suppression for general $N$ (App.~\ref{App:discrete}).

\section{N-suppression}\label{Sec:N-suppression}

Twin Higgs models~\cite{Chacko:2005pe,TH2} use a $Z_2$ exchange symmetry between the SM and the twin sector
to efficiently soften the Higgs potential. This $Z_2$ could for example originate from the Higgs trigonometric parity (TP)~\cite{Csaki:2017jby} naturally present in symmetric coset spaces.
However the Higgs VEV in the presence of this $Z_2$ symmetry is too large for realistic EWSB. Hence one usually assumes that the $Z_2$ symmetry is broken either in the top or in the gauge sector. This will allow a realistic Higgs VEV at the price of  reintroducing at least 10 percent fine tuning in the Higgs potential.

Here we propose a new mechanism to obtain natural and realistic EWSB in twin Higgs models without the need for breaking the $Z_2$ symmetry at all. The novel ingredient in our model is that the effective Yukawa terms only depend on $\sin(Nh/f)$ and $\cos(Nh/f)$, rather than   $\sin(h/f)$ and $\cos(h/f)$ as it is usually assumed. In this case the $Z_2$ Twin Higgs exchange symmetry  will be implemented in the top sector as a TP corresponding to the exchange symmetry between $\sin(Nh/f)$ and $\cos(Nh/f)$ where $N$ is an odd integer. The beauty of this idea is that while the full $Z_2$ can remain exactly unbroken, the structure of the Higgs potential can significantly differ from the traditional case and provide a realistic EWSB minimum without the need of breaking the $Z_2$ symmetry at all. In particular this setup with the  ``N-trigonometric parity'' will significantly reduce the magnitude of the Yukawa coupling, which will result in the suppression of the Higgs quadratic term while leaving the quartic term essentially unchanged.   In the rest of this section we will give a more detailed overview of this mechanism, while  in the next section we present the general mechanism that will actually generate the $\sin(Nh/f)$($\cos(Nh/f)$) dependent effective Yukawa terms.

In ordinary twin Higgs models, a twin top $\tilde{t}$ (which is a SM singlet but a triplet under twin QCD $SU(3)_c^\prime$) is introduced to implement the $\tilde{t} \leftrightarrow t$ $Z_2$ exchange symmetry. The presence of this exchange symmetry will ensure that the trigonometric parity transformation
\bea \label{eq:TP}
 \frac{h}{f} \to -\frac{h}{f} +\frac{\pi}{2} , \ \ \  \sin\frac{h}{f} \leftrightarrow \cos\frac{h}{f}\,
\eea
is left unbroken by the top sector.
We can impose in addition maximal symmetry \cite{Csaki:2017cep}, which will imply that only the
Yukawa couplings will be Higgs dependent in the effective Lagrangian of the top sector (after integrating out the heavy vectorlike fermions). Our new ingredient will be the assumption that the
effective Yukawa couplings actually depend on $\sin(Nh/f)$ and $\cos(Nh/f)$ (rather than $\sin h/f$ and $\cos h/f$ as usual). We will motivate this assumption in the next section, for now we will just explore what the consequences of a Yukawa coupling of the form
\bea\label{eq:eff_yukawa}
\mathcal{L}_{t}^\text{Yuk.}=y_t^\prime f\bar{t}_L t_R \sin(\frac{Nh}{f}) +\tilde{y}_t^\prime f \bar{\tilde{t}}_L \tilde{t}_R \cos(\frac{Nh}{f})+h.c.
\eea
would be. The effect of the $Z_2$ TP  in Eq.(\ref{eq:TP}) on these trigonometric functions  (if $N$ is an odd integer) is
\begin{align}\label{eq:Ntrigno_parity}
&  s_{Nh}  \leftrightarrow c_{Nh},\;\: \text{for}\: N=4n+1\nonumber\\
 & s_{Nh}  \leftrightarrow -c_{Nh},\; \text{for}\: N=4n+3,
\end{align}
where $s_{Nh}\equiv\sin(\frac{Nh}{f}) (c_{Nh} \equiv \cos(\frac{Nh}{f}))$. Hence the $Z_2$ exchange symmetry is maintained by the Yukawa couplings if  $y'_t =\pm\tilde{y}'_t$ with the sign determined by $N\mod 4$ as in Eq.(\ref{eq:Ntrigno_parity}). After integrating out the fermions the$\mathcal{O}(y_t^{\prime2})$ terms in the Higgs potential will cancel due to this ``N-trigonometric parity'' ($s_{Nh}  \leftrightarrow \pm c_{Nh}$), and the leading order contributions will be at $\mathcal{O}(y_t^{\prime4})$ of the form
\bea\label{eq:fermion_Pot}
V_f(h) = c_t \frac{N_c y_t^{\prime 4} f^4}{(4\pi)^2}\Big(s_{Nh}^4 + c_{Nh}^4 \Big),
\eea
where $N_c=3$ is the number of (ordinary and twin) colors, $c_t \simeq \log( \Lambda_f^2 /m_t^2 )$ and $\Lambda_f < 4\pi f$ is the cut-off scale in the fermions sector. It is then easy to find the VEV corresponding to the minimum of the Higgs potential from the top-twin-top sector $V_f$:
\bea \label{eq:natural_VEV}
\frac{\langle h\rangle}{f} = \frac{\pi}{4 N}\ .
\eea
The natural value of $\langle h \rangle /f$  is suppressed by $N$ and it can be much smaller than 1 if $N\geqslant 3$! The origin of the naturally small Higgs VEV is due to the strong suppression of the Higgs quadratic term in the Higgs potential (while the quartic has no additional suppression).
  In order to explicitly see this, we first need to clarify the relation between $y'_t$ and the SM Yukawa coupling $y_t$. The top acquires a mass after electroweak symmetry breaking (EWSB), $\langle h \rangle =v_{SM}$,
\bea
m_t = y_t^\prime f  s_{Nh}\equiv y_t v_{SM}.
\eea
Using the above expression, we find that the deviation of top Yukawa coupling from its SM value is
\bea
\frac{y_{tt h}^\prime }{y_{t} } \thickapprox N \sqrt{\xi} \cot ( N \sqrt{\xi}),
\eea
where $y_{tth}^\prime$ is the physical top Yukawa coupling in our model while $\xi\equiv s_h^2\approx (v_{SM}/f)^2$ is the usual parameter that measures the hierarchy between $v_{SM}$ and $f$. Experimentally we know that $\frac{y_{tt h}^\prime }{y_{t} }$ should be close to 1 (deviations of order 10 percent are still possible), hence we need to ensure $N \sqrt{\xi} \ll 1$.
While the natural value of $\xi$ from the top sector alone from Eq.~(\ref{eq:natural_VEV}) would not satisfy this relation, we will see in a moment that adding the gauge contribution to the potential will automatically rectify this (at least for not too large values of $N$) and ensure that the relation  $N \sqrt{\xi} \ll 1$ be satisfied.  This will also imply  $N h/f \ll 1$, and thus the approximate expression for the top mass is $m_t \approx N y_t^\prime v_{SM}$. Hence we finally find that $y_t^\prime$ is suppressed by $N$ compared to the SM Yukawa coupling
\bea \label{eq:yukawa}
y_t^\prime \approx  \frac{y_t}{N}.
\eea
Once we established the expression for $y_t'$ and that  $N h/f \ll 1$ we can expand the Higgs potential of the top sector $V_f $ in term of $s_h$ up to $\mathcal{O}(s_h^4)$ to find
\bea
V_f(h) \approx  c_t \frac{2 N_c y_t^{4} f^4}{(4\pi)^2}\left(-\frac{1}{N^2} s_{h}^2 + \frac{4N^2-1}{3N^2}s_{h}^4 \right).
\eea
As promised, we find that the $s_h^2$ term is suppressed by $N^2$  while $s_h^4$ term is almost unchanged compared to the $N=1$ case of the ordinary twin Higgs. This suppression is the reasons why the Higgs VEV is naturally small. However we still need to add the contributions of the gauge sector - we just saw above that those are important as well. In the gauge sector we assume that the implementation of the $Z_2$ TP  is the same as in the ordinary twin Higgs model and we summarize it in App.~\ref{App:gauge}. The gauge contribution is at $\mathcal{O}(g^4)$ and can be parametrized as
\bea
V_g(h) = -c_g\frac{9g^4 f^4 }{64 (4\pi)^2}\Big(s_h^4 + c_h^4 \Big),
\eea
where $c_g \approx \log (\Lambda_g^2 /m_W^2 )$ and $\Lambda_g$ is the cut-off scale in the gauge sector. The full potential can thus be parameterized as
\bea
V(h) \approx -(\frac{\gamma_f^\prime}{N^2} -\gamma_g^\prime) s_h^2 + \frac{4}{3}\gamma_f^\prime s_h^4,
\eea
where we defined $\gamma_f^\prime \equiv2 c_t  N_c y_t^{4} f^4 /(4\pi)^2$ and $\gamma_g^\prime \equiv9c_g g^4 f^4 / (32(4\pi)^2)$, and neglected the $s_h^4$ term from gauge sector because $\gamma_g^\prime \ll \gamma_f^\prime$. We can easily find the Higgs VEV and mass for this potential
\bea\label{eq:xiHiggs}
\xi \approx \frac{\frac{\gamma_f^\prime}{N^2} -\gamma_g^\prime}{8 \gamma_f^\prime/3}, \quad m_h^2 \approx \frac{32\gamma_f^\prime \xi (1-\xi)}{3f^2}.
\eea
Since the $s_h^4$ term is almost the same as in the ordinary twin Higgs model (see for example\cite{Csaki:2017jby}) and insensitive to the cut-off scale, one can take $\Lambda_f$ very high while still keeping a naturally light Higgs.

The only tuning in this model comes from the necessary cancellation in the $s_h^2$ between the top and the gauge sectors, which is needed to ensure $N \sqrt{\xi} \ll  1$. In generic composite Higgs models $\gamma_g^\prime$ is much smaller than $\gamma_f^\prime$ so one might worry this is a serious source of tuning. The beauty of our mechanism is that the N-suppression will reduce the effective $\gamma_f^\prime$ such that it's magnitude can be automatically of the same order as $\gamma_g^\prime$ and the partial cancelation can be natural without much tuning, as long as $N$ is not too large, for example $N=3$. Note that this cancelation here is very different from the usual sources of tuning in Twin Higgs models. Generic Twin Higgs models introduce an explicit $Z_2$ breaking term in order to obtain the correct EWSB minimum, which will reintroduce a quadratic dependence to the cutoff scale $\Lambda_g$ or $\Lambda_f$ resulting in some tuning 
in these models. For example, if the gauge sector breaks the $Z_2$ symmetry, the tuning is around $1\%$ for gauge boson partner masses lighter than $9$ TeV (with $\xi=0.01$ fixed) but increases linearly with the square of partner's mass for masses heavier than around 9 TeV \cite{Csaki:2017jby}. Using the N-suppression mechanism proposed here will allow us to keep the $Z_2$ TP completely unbroken and still obtain a realistic EWSB minimum for the Higgs potential, hence the cutoff dependence is never reintroduced here, and the tuning will be much smaller than in existing TH models. We will present a detailed discussion of the tuning in Sec. \ref{sec:tuning}.

\section{Realization of N-suppression}\label{sec:Fermion_sector}

In this section we focus on how to generate a $\sin(Nh/f)$ ($\cos(Nh/f)$) dependent effective Yukawa coupling to realize N-suppression based on the maximally symmetric twin Higgs model. The minimal coset space to realize the twin Higgs mechanism in the gauge sector is $SO(8)/SO(7)$. Our setup for the gauge sector will be the same as the usual $SO(8)/SO(7)$ twin Higgs model and we summarize it in App.~\ref{App:gauge}. In this coset space, the physical Higgs is identified with the $SO(2)$ rotation angle between $4^{th}$ and $8^{th}$ directions of $SO(8)$. In unitary gauge, the pNGB matrix is
\bea
U=e^{i\frac{\sqrt{2}}{f}h^{\hat{a}}T^{\hat{a}}}=\left( \begin{array}{cccc}
 \mathds{1}_3   & 0 & 0 & 0 \\
0 & \cos \frac{h}{f}  & 0 & \sin \frac{h}{f} \\
0 & 0 &  \mathds{1}_3 & 0 \\
0 & -\sin \frac{h}{f} & 0 & \cos \frac{h}{f}
\end{array} \right),
\eea
where $T^{\hat{a}}$'s are the broken generators. Since $U$ is just the $SO(8)$ element and $h/f$ is the rotation angle we can find the key point to get $\sin(Nh/f)$ ($\cos(Nh/f)$) is that we must insert $N$ pNGB matrix $U$ in effective Yukawa term: $U^N=U(h\rightarrow Nh)$.

\subsection{General Mechanism to Generate $U^N$ Yukawa Term}

To ensure that the top Yukawa interaction is the only Higgs dependent term in the low-energy effective Lagrangian, we will make use of maximal symmetry. As explained in~\cite{Csaki:2018zzf}, maximal symmetry is a global $SO(8)$ symmetry in the massive fermion sector, which is a subgroup of the chiral symmetries of the fermions. This leftover global symmetry guarantees that the Higgs shift symmetry remains unbroken by effective kinetic terms and is only broken by the effective top Yukawa terms.  Hence only the effective top Yukawa term will be  Higgs dependent.
In addition we will assume that the shift symmetry is collectively broken by the Yukawa couplings in the top sector containing additional massive fermions in a special ``chain form''. The top mass will be collectively generated through these mixing terms and will give rise to the desired $U^N$ dependence. Next we will discuss how to realize this special collectively generated top mass using maximal symmetry.

The UV completed $SO(8)/SO(7)$ CHM can be constructed based on an extra dimension or its deconstructed  version.   On the elementary site, the $SU(2)_L\times U(1)_Y$ as well as $SU(2)_L^\prime\times U(1)_Y^\prime$ subgroup of global symmetry $SO(8)_{el}$ are gauged corresponding to the usual and twin EW gauge symmetries. The SM and twin SM fermions live on the elementary site and we embed the left hand doublets ($q_L,\,\tilde{q}_L$) in the fundamental representation of $SO(8)_{el}$ as
\bea
\Psi_{q_L} = \frac{1}{\sqrt{2}} \left( \begin{array}{c}
i b_L  \\
b_L  \\
i t_L  \\
- t_L  \\
0 \\
0\\
0 \\
0
\end{array}  \right), \quad \Psi_{\tilde{ q}_L} =\frac{1}{\sqrt{2}} \left( \begin{array}{c}
0 \\
0\\
0 \\
0\\
i \tilde{b}_L  \\
\tilde{b}_L  \\
i \tilde{t}_L \\
- \tilde{t}_L
\end{array}  \right),
\eea
while the right handed $t_R$ and $\tilde{t}_R$ are treated as $SO(8)_{el}$ singlets. After we integrate out all the composite modes, there will only be one elementary global $SO(8)$ symmetry left in the effective Lagrangian. The pNGB fields will show up as the linearly realized sigma field $\Sigma'=UVU^\dag$  in  symmetric representation of $SO(8)$ or $\mathcal{H}=U\mathcal{V}$ which is in the fundamental representation of $SO(8)$, with the transformations under the  global symmetry acting as $\Sigma'\rightarrow g\Sigma'g^\dag$  and $\mathcal{H}\rightarrow g\mathcal{H}$. Where the Higgs parity operator $V=\text{diag}(1,1,1,1,1,1,1,-1)$ and $\mathcal{V}=(0,0,0,0,0,0,0,1)$ are separately the symmetric and vector-like VEV that break $SO(8)$ to $SO(7)$. Both can be assumed to originate from some scalars VEVs that break $SO(8)$ to $SO(7)$ at some UV scale.

\begin{figure}
  \centering
  \includegraphics[width=9cm]{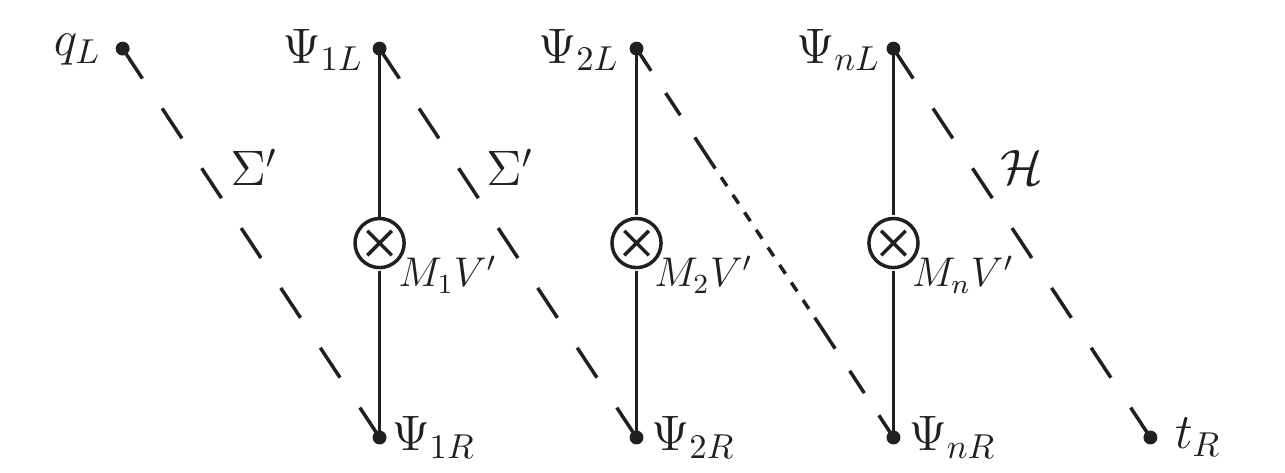}\\
  \caption{Yukawa couplings between the fermion multiplets that collectively break the pNGB shift symmetry in a special ``chain form". There is a similar construction for the twin sector.}\label{fig:Yukawa}
\end{figure}

In order to realize the novel structure of interactions including $U^N$ in the low energy effective Lagrangian, we must introduce the additional Dirac fermions $\Psi_i$ ($\tilde{\Psi}_i$) forming complete vector representations of $SO(8)$ and $SU(3)_c$ ($SU(3)_c^\prime$) at the elementary site. These will mix with the SM and twin tops. The key ingredient for obtaining a $U^N$ dependent top Yukawa is the particular pattern of collective breaking of the Higgs shift symmetry
 by the Yukawa couplings of $\Psi_i$ ($\tilde{\Psi}_i$) shown in Fig.~\ref{fig:Yukawa}, where we introduce $n$ multiplets ($N=2n+1$). In addition, we have to ensure the emergence
of maximal symmetry which will protect the effective kinetic terms from Higgs dependent corrections in the low energy effective theory. To achieve this, the masses of $\Psi_i$ ($\tilde{\Psi}_i$) should be twisted by $V'=\text{diag}(1,1,1,1,-1,-1,-1,-1)$ to explicitly break the chiral global symmetry $SO(8)_L\times SO(8)_R$ of each of these Dirac fermions to the $SO(8)_{V'}$ subgroup identified with maximal symmetry. Note that the choice of $V'$ as the mass term preserves the EW and twin EW gauge symmetries while at the same time ensures the proper breaking of the global symmetries.  With this symmetry breaking pattern  the low-energy effective Lagrangian will have  the form (after integrating out the massive Dirac fermions)
\begin{equation}
  \mathcal{L}_t^\text{Yuk.}\sim M^t_1 \bar{\Psi}_{q_L} \Sigma  t_R+\tilde{M}^t_1 \bar{\Psi}_{\tilde{q}_L} \Sigma  \tilde{t}_R+h.c.,
\end{equation}
where $\Sigma=\Sigma'V'\Sigma'\cdot\cdot\cdot V'\mathcal{H}$. Notice that $V'$ has the same  effect as the Higgs parity $V$ acting on the pNGB Higgs matrix, $UV=VU^\dagger$ and $UV'=V'U^\dagger$. Thus we can easily show that
\begin{equation}
  \Sigma=U^{2n+1}\mathcal{V}=\left( \begin{array}{c}
 {\bf 0} \\
 s_{Nh}\\
 {\bf 0}  \\
 c_{Nh}
\end{array} \right),
\end{equation}
where $N=2n+1$ is an odd integer (in agreement with our earlier claim in Sec.\ref{Sec:N-suppression} that the N-suppression mechanism requires an odd $N$). Thus we have shown how to explicitly realize the effective Lagrangian in Eq.(\ref{eq:eff_yukawa}). In general one would expect that all Dirac fermion multiplets could arbitrarily mix with each other because they all transform under the same global symmetry $SO(8)_{el}$, which would result in additional $U^i$ ($i<n$) dependent terms in the effective Lagrangian, potentially ruining our N-suppression mechanism. However we can impose a $Z_2$ parity which will enforce the mixing pattern displayed in Fig.~\ref{fig:Yukawa}. The details of this $Z_2$ parity are shown in App.~\ref{App:discrete}.    Notice that in principle we could have also chosen the unit matrix instead of the twisted $V'$ to realize maximal symmetry, which would also preserve the gauge symmetries. This case however would not give us the hoped-for $U^N$ dependence in the effective action,  since using $\Sigma^\prime \Sigma^\prime =1$ in the case of the unit mass matrix the $\Sigma'$ dependence would drop out.

\subsection{Realistic Model With $N=3$}
\label{section:2}

In this subsection we present the detailed discussion of the $N=3$ case,  where only one elementary multiplet needs to be introduced. While $N=3$ is clearly the simplest model implementing our ideas, it turns out that larger values of $N$ will not work well at least within our simple implementation of the gauge sector. The reason is that the cutoff scale of the gauge sector (and hence the mass of the gauge partner $m_\rho$) will rapidly decrease with the increase of $N$.  The reason behind is that  the overall $\gamma= \gamma_f^\prime/N^2-\gamma_g^\prime$ and $\gamma_f^\prime$ are almost fixed positive values set by the Higgs mass (see eq.~(\ref{eq:xiHiggs})). Hence if $N$ increases
$\gamma_g^\prime$ must decrease if we want to keep $\gamma$ fixed. However $\gamma_{g}^\prime$ is only lograithmically sensitive to $m_\rho$, hence the increase in $N$ will require a drastic decrease in $m_\rho$, which will quickly become lighter than the experimental bound. This argument shows that $N=3$ needs to be chose for a realistic model. We will see that for $N=3$ the Higgs potential is indeed natural. If one were to increase $N$, there will be an increasing amount of tuning needed to ensure that $m_\rho$ is sufficiently heavy.  Notice that $N$ is an integer and its value is expected to be determined by the choice of UV completion.
Choosing a particular value of $N$ is equivalent to choosing a particular UV completion that can generate Higgs potential suppressed by a particular factor $N^2$.  Hence the choice of $N=3$ is not expected to be increasing the tuning of the model. 

   As we explained before, on the elementary site we introduce the Dirac fermion $\Psi$ and $\tilde{\Psi}$ which are triplets under QCD/twin QCD as well as forming complete vector  representations of the global $SO(8)$ symmetry. We assume that the mass terms of $\Psi$ and $\tilde{\Psi}$ are twisted by $V'$ which breaks the chiral global symmetry $SO(8)_L\times SO(8)_R$ to the $SO(8)_{V\prime}$ maximal symmetry so that the effective Yukawa coupling is the only Higgs dependent term in the effective Lagrangian. In addition, we should also preserve the Higgs TP, which is the combination of a $\pi/2$ rotation in the Higgs direction with the Higgs parity $V$. The combined transformation has the form
\bea
P_1^h=\left(\begin{array}{cccc}
 \mathds{1}_3   &  &  &  \\
 &   &  &-1\\
 &  &   \mathds{1}_3 &  \\
 & -1&  &
\end{array} \right),
\eea
and $U$ transforms under TP as $U \to P_1^h U V^\dagger =U(s_h \leftrightarrow c_h)$. In order to preserve TP, the fermion interactions should be invariant under the $Z_2$ exchange symmetry between the SM sector and the twin sector defined as
\begin{align}
\Psi_{q_L} &\leftrightarrow P \tilde{\Psi}_{q_L},\;t_R\leftrightarrow\tilde{t}_R,\;\Psi\leftrightarrow P\tilde{\Psi},\nonumber\\
 U &\to P U V^\dagger P_0=U(s_h \leftrightarrow c_h),
\end{align}
where $P$ is an operator implementing the exchange of the top and the twin top of the form
\bea
P=P_0 P_1^h = \left( \begin{array}{cc}
0&\mathds{1}_4   \\
\mathds{1}_4 & 0 \\
\end{array}  \right) ,\, P_0=\left(
                                          \begin{array}{cccc}
                                             &  & \mathds{1}_3 &  \\
                                             & -1 &  &  \\
                                            \mathds{1}_3 &  &  &  \\
                                             &  &  & -1 \\
                                          \end{array}
                                        \right) \ .
\eea
$P_0$ acts trivially on the Higgs pNGB matrix and commutes with $P_1^h$, $U$ and $V$. Using the Yukawa couplings shown in Fig.~\ref{fig:Yukawa}, we can easily write down the Lagrangian invariant under the $Z_2$ exchange symmetry and the $SO(8)_{V^\prime}$ maximal symmetry:
\begin{align}\label{eq:Lagrangian}
\mathcal{L}_f&= \bar{q}_L i \slashed D q_L +\bar{t}_R i\slashed D t_R + \bar{ \Psi} i \slashed D \Psi       \nonumber \\
&-  \epsilon_{L} f  \bar{\Psi }_{q_L} \Sigma' \Psi_{R}- M \bar{\Psi}_{R} V^\prime \Psi_{L}- \epsilon_{R}f \bar{ \Psi}_{L} \mathcal{H}t_R
\nonumber \\
&+ \Big( q_L , t_R , \Psi,M \to  \tilde{q}_L , \tilde{t}_R , \tilde{\Psi},-M   \Big) + h.c.,
\end{align}
where $D_\mu=\partial_\mu -i( g W_\mu ^a \tau_L^a +g^\prime  B_\mu \tau_{R}^3) - i (\tilde{g} \tilde{W}_\mu ^a \tilde{\tau}_L^a +\tilde{g}^\prime  \tilde{B}_\mu \tilde{\tau}_{R}^3 )$ where $\tau_{L,R}^a$ and $\tilde{\tau}_{L,R}^{a}$ are the $SU(2)_{L,R}$ and their mirror $SU(2)_{L,R}^\prime$ generators.

As in~\cite{Csaki:2018zzf}, the twisted mass terms of the elementary top partners can be assumed to originate from their Yukawa couplings to a scalar $\Phi$ in the symmetric representation of the  global $SO(8)$ which acquires a VEV $V^\prime$ at some higher energy scale to break the global symmetry $SO(8)$ to $SO(4)_1 \times SO(4)_2$. We emphasize again that $V^\prime$ is the only choice (besides the identity) that preserves the gauge symmetries. The NGBs from this scalar multiplet can get sufficiently large mass corrections from the gauge loop or also from the tree level term $\text{Tr}[V^\prime \Phi]^2$, which explicitly breaks $SO(8)$ to $SO(4)_1 \times SO(4)_2$.   In order to obtain the correct Yukawa couplings, we impose three different $Z_2$ symmetries: one each for the fields  $\Psi_{q_L}$, $\Psi$ and $t_R$ and assume they have odd parity under their corresponding $Z_2$ symmetry, while even under the other two. Hence $\Psi_{q_L}$ has parities $(-,+,+)$, $\Psi$ has $(+,-,+)$ while  $t_R$ has $(+,+,-)$.  If we also assume that the pNGB fields $\Sigma'$ and $\mathcal{H}$ have the parities $(-,-,+)$ and $(+,-,-)$ under the $Z_2\times Z_2\times Z_2$ symmetry, one can easily check that the terms  in  (\ref{eq:Lagrangian}) are the only ones allowed by the $Z_2^3$ symmetry. The direct mixing of $q_L$ and $\Psi$ as well as the Yukawa coupling of $q_L$ and $t_R$ are both $Z_2$ odd and thus forbidden. The parities of the pNGB fields can be assumed to originate from the scalars at the last site in the $SO(8)/SO(7)$ moose diagram whose VEV $V$ or $\mathcal{V}$ break the gauged $SO(8)$ at the last site to $SO(7)$, similar to the set up considered in~\cite{Csaki:2018zzf}. This $Z_2^3$ symmetry can be easily generalized to a $Z_2^N$ symmetry for the case of the general $N$ case depicted in Fig.~\ref{fig:Yukawa}. For a more detailed discussion see App.~\ref{App:discrete}.

After integrating out the elementary partners, the effective Lagrangian in momentum space has the form (we neglect the gauge covariant kinetic terms for simplicity)
\begin{align}
\mathcal{L}_{t}^{\text{eff}}&= \bar{\Psi}_{q_L}\slashed p \Pi_0 ^q \Psi_{q_L}+\bar{t}_R \slashed p \Pi_0 ^t t_R  - M_1 ^t \bar{\Psi}_{q_L} \Sigma  t_R  \nonumber \\
&+  \Big( q_L , t_R,M_1^t  \to  \tilde{q}_L , \tilde{t}_R,-M_1^t   \Big)+  h.c.,
\end{align}
where
\bea
\Sigma\equiv\Sigma'  V^\prime U \mathcal{V} = U^3 \mathcal{V} =  \left( \begin{array}{c}
 {\bf 0} \\
 s_{3h}\\
 {\bf 0}  \\
 c_{3h}
\end{array} \right)
\eea
and the form factors are
\begin{equation}
  \Pi_0 ^{q,t} =1-\frac{\epsilon_{L,R} ^2 f^2}{p^2 -M ^2},\quad M_1 ^t = \frac{\epsilon_L \epsilon_R f^2 M}{p^2 -M ^2 }.
\end{equation}
We see that as expected maximal symmetry forbids the Higgs dependent terms in the effective kinetic terms and the Yukawa terms are proportional to $s_{3h}$ or $c_{3h}$. Since the form factors in the top and twin top sectors are equal, the $Z_2$ exchange symmetry can be easily checked by the transformation of $\Sigma$ under $Z_2$ exchange operator $P$,
\bea
\Sigma \to -P \Sigma =\Sigma(s_{3h} \leftrightarrow -c_{3h} ),
\eea
where the minus sign arises from the commutator $[P,V']=-1$. Hence the effective Lagrangian is invariant under the exchange symmetry between the top and the twin top and Higgs TP, $s_{3h}\leftrightarrow -c_{3h}$. Thus the  Higgs potential at one-loop level must also be invariant under this TP. The $Z_2$ symmetry can be explicitly seen from the expansion of effective Lagrangian in terms of $t$ and $\tilde{t}$,
\begin{align}
\mathcal{L}_{t}^{\text{eff}}&= \bar{t}_L \slashed p \Pi_0 ^q t_L + \bar{t}_R \slashed p \Pi_0 ^t t_R  +\bar{b}_L \slashed p \Pi_0 ^q b_L  \nonumber \\
&+ \bar{\tilde{t}}_L \slashed p \Pi_0 ^q \tilde{t}_L + \bar{ \tilde{t} }_R \slashed p \Pi_0 ^q \tilde{t}_R +\bar{\tilde{b} }_L \slashed p \Pi_0^q \tilde{b}_L \nonumber \\
&+ \frac{M_1 ^t }{\sqrt{2}} \bar{t}_L t_R s_{3h}-\frac{M_1 ^t }{\sqrt{2}} \bar{\tilde{t} }_L \tilde{t}_R c_{3h} +h.c.
\end{align}
The top mass can be extracted from the above Lagrangian,
\bea \label{eq:mt}
m_t =\frac{\epsilon_{L} \epsilon_{R} f^2 M  }{\sqrt{2}M_L M_R} s_{3h}  \equiv y_t ^\prime f s_{3h},
\eea
where $M_{L,R} = \sqrt{\epsilon_{L,R} ^2 f^2 +M ^2}$ are the top partner masses. As explained before, for realistic VEVs $3\langle h \rangle/f \ll 1$, the relation between $y_t^\prime$ and SM top Yukawa is $y_t\approx3y'_t$. The Higgs potential from top and twin top loops is given by
\begin{align}
V_f &= -2N_c   \int \frac{d^4p_E}{(2\pi)^4} \text{log}\Big[1 + \frac{| M_1^t |^2 }{2 p_E^2\Pi_0^q \Pi_0^t} s_{3h}^2  \Big] \nonumber \\
&+ (s_{3h} \to -c_{3h}),
\end{align}
where $N_c =3$ is the number of QCD colors. The Higgs potential at $\mathcal{O}(y_t^{\prime2})$ is proportional to $s_{3h}^2+c_{3h}^2$ and cancelled by Higgs TP. Thus the leading order Higgs potential is at $\mathcal{O}(y_t^{\prime4})$ and can be parameterized as
\begin{equation}
  V_f\approx \gamma_f(s_{3h}^4+c_{3h}^4),
\end{equation}
where
\begin{equation}\label{eq:gamma_f}
  \gamma_f=\frac{N_c}{(4\pi)^2}  \int^{\infty}_{m_t^2} dp_E^2 p_E^2 \Big( \frac{|M_1^{t}|^2}{2p_E^2 \Pi_0^q \Pi_0^t}  \Big)^2.
\end{equation}

\section{EWSB and Fine Tuning\label{sec:tuning}}

In this section we discuss the Higgs potential of our $N=3$ benchmark model and the resulting pattern of EWSB. We find that there is essentially no tuning needed to achieve realistic EWSB with  heavy top and gauge partners, thanks to our novel N-suppression mechanism. We want to also emphasize again that while our N-suppression is based on a twin-Higgs like scenario in the presence of maximal symmetry, it has an important difference from the ordinary twin Higgs models. In the ordinary twin Higgs models one always needs a source of explicit $Z_2$ breaking in order to produce realistic EWSB. However here the softening of the Higgs potential due to the N-suppression will allow us to build a model where the $Z_2$ symmetry is exactly preserved both in the fermion and gauge sectors~\footnote{Another interesting way to preserve $Z_2$ symmetry in twin-Higgs model is by introducing an extra Higgs quartic term in the potential \protect\cite{Csaki:2019coc}.}.

\begin{figure}
\begin{center}
\includegraphics[width=0.49\columnwidth]
{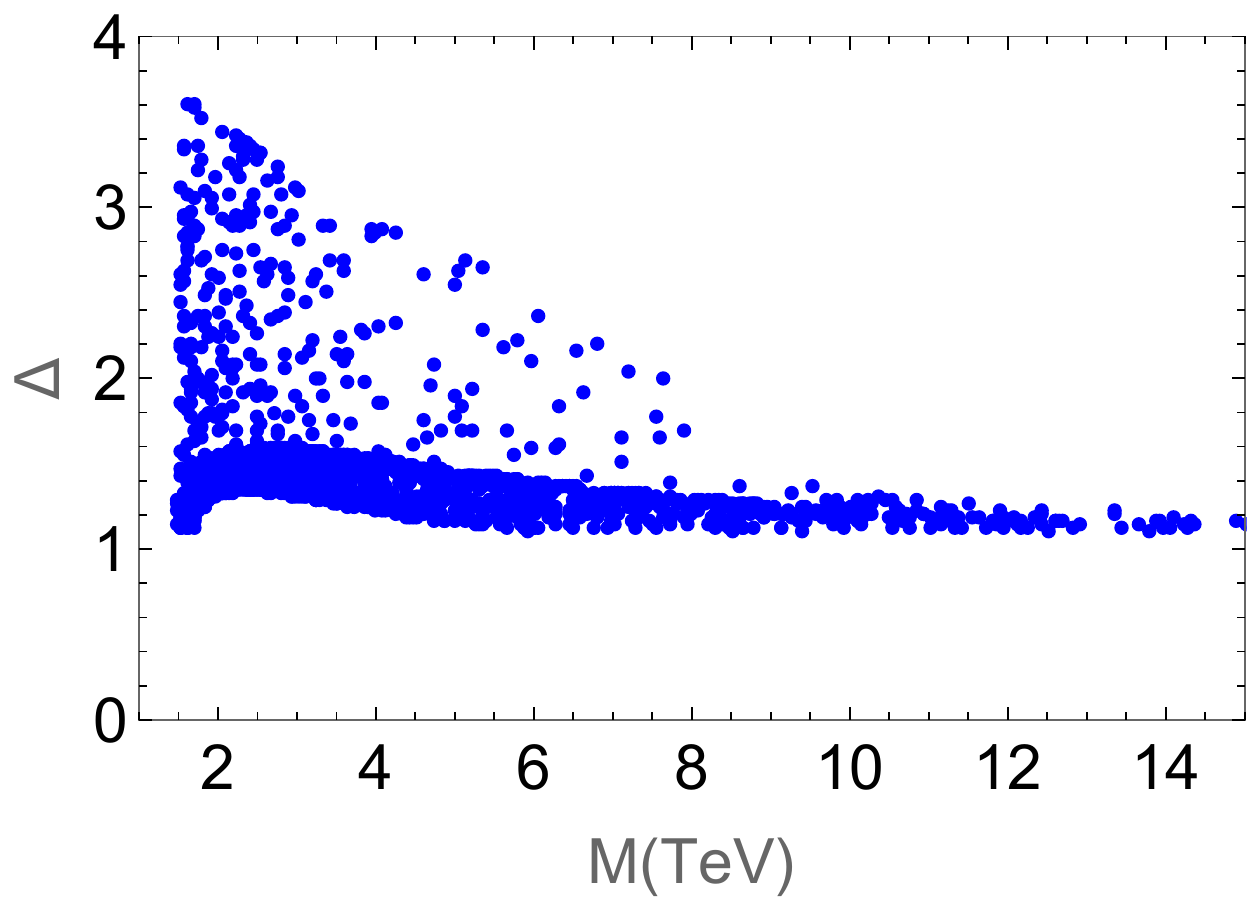}
\includegraphics[width=0.49\columnwidth]
{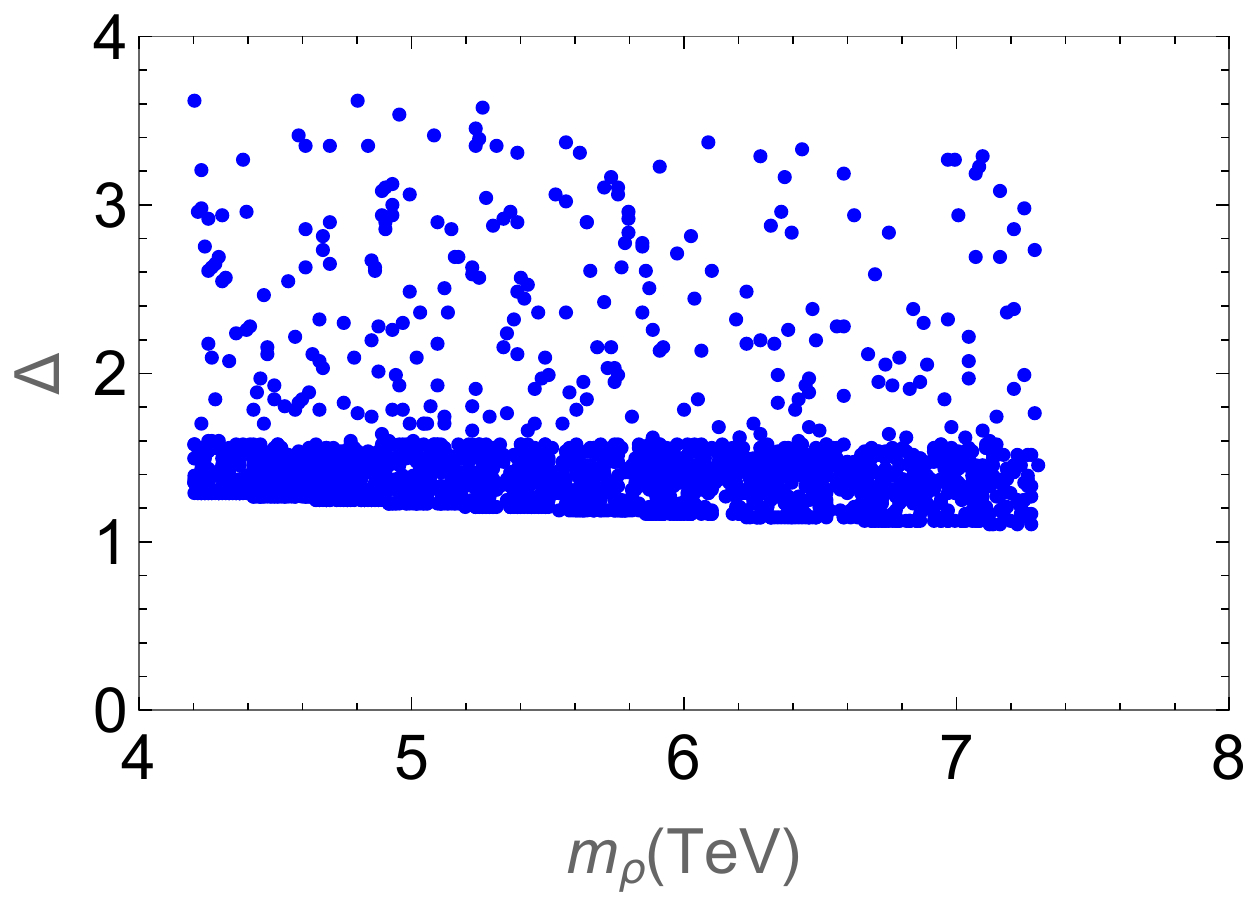}
\includegraphics[width=0.90\columnwidth]
{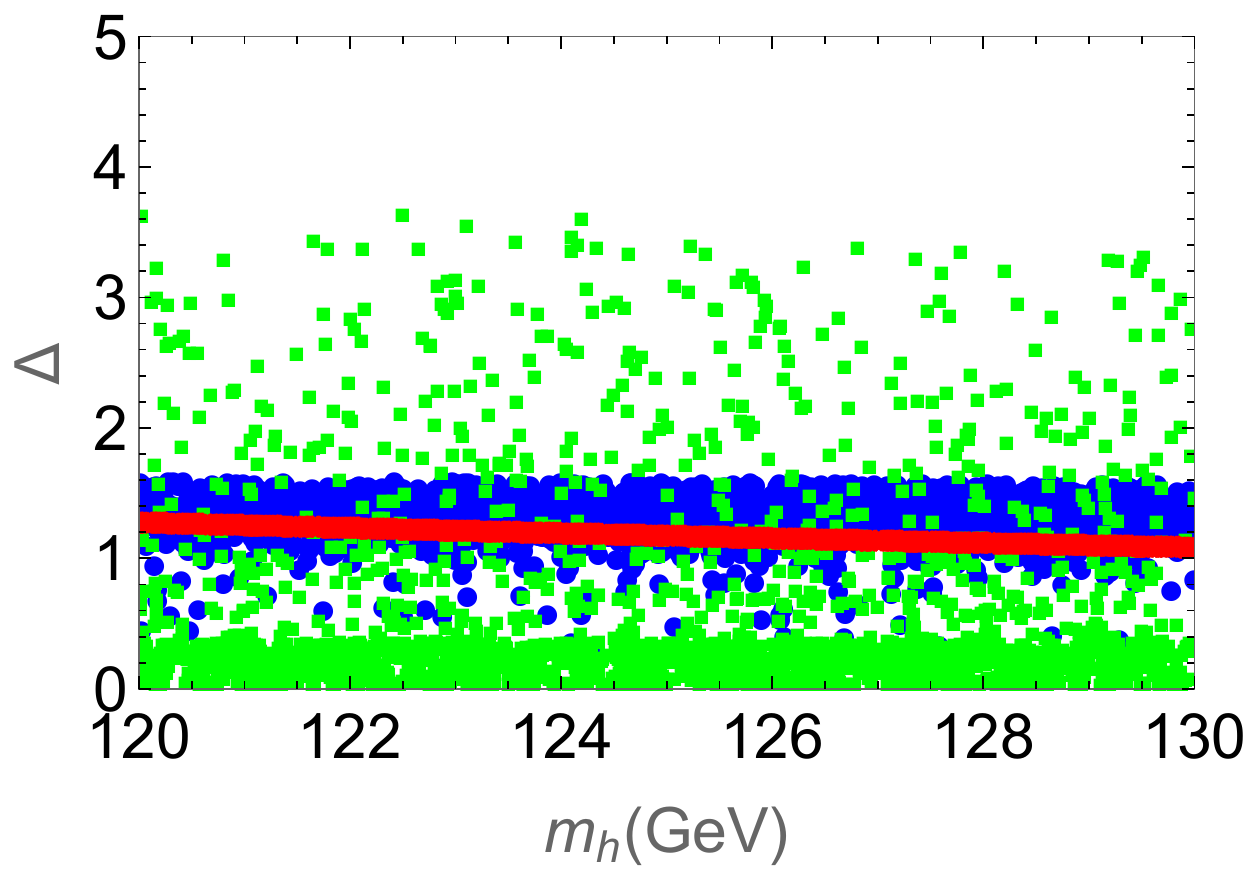}
\end{center}
\caption{Scatter plot in N-suppression twin Higgs model $SO(8)/SO(7)$ for $\xi=0.01$. The range of the parameters is taken as follows: $m_t \in [140, 170]$ GeV, $M >1.5$ TeV and $m_\rho >2.5$ TeV. Top: in left panel fine tuning $\Delta$ as a function of $M$ and in right panel as a function of $m_\rho$ for $m_h \in [120,130]$ GeV. Bottom: Scatter plot of tuning $\Delta_i$ for various input parameters $x_i$, $M$(blue), $\epsilon_L$(green) and $m_\rho$(red), as a function of $m_h$.}
\label{fig:mh_mf_001}
\end{figure}

Based on our discussion of the fermion sector in Sec.\ref{sec:Fermion_sector} as well as the gauge sector in App.~\ref{App:gauge}, the leading contributions to the Higgs potential in our $N=3$ model can be parametrized as
\begin{equation}
  V(h)=\gamma_f(s_{3h}^4+c_{3h}^4)-\gamma_g(s_h^4+c_h^4),
\end{equation}
where $\gamma_{g,f}>0$ and their actual expressions can be easily estimated from Eqs.~(\ref{eq:gamma_f}) and (\ref{eq:gamma_g}):
\begin{equation}\label{eq:gamma_fg}
  \gamma_f\simeq c_f^\prime\frac{N_cy_t^{\prime4}f^4}{(4\pi)^2}\ln\frac{M_f^2}{m_t^2},\,\gamma_g\approx c_g^\prime\frac{9f^4g^4}{1024\pi^2}\Big(\ln\frac{32m_\rho^2}{m_W^2}-5\Big),
\end{equation}
where $M_f$ is a typical top partner mass scale and $m_\rho$ is the gauge partner mass, $c_{f,g}^\prime$ are $\mathcal{O}(1)$ numerical parameters, which can be calculable in our  model. In our  numerical expressions of the tuning presented below we always use their exact expressions in Eqs.~(\ref{eq:gamma_f}) and (\ref{eq:gamma_g}), while the above expressions are very useful to give an order of magnitude estimate for the tuning. When  calculating the minimum of the potential, we find that for a realistic EWSB one needs  $9\gamma_f-\gamma_g>0$. Then the minimum of this potential is at
\bea\label{eq:xi}
\xi \equiv \langle s_h^2 \rangle =\frac{1}{2} -\frac{1}{2} \sqrt{\frac{1}{2} + \frac{1}{4}\sqrt{1+\frac{\gamma_g}{3 \gamma_f}  } },
\eea
and after EWSB the physical Higgs mass is given by
\bea\label{eq:Higgs_mass}
m_h^2 =\frac{4(9\gamma_f -\gamma_g)\sqrt{1+ \frac{\gamma_g}{3\gamma_f}} }{f^2}.
\eea
Because of the $N$ suppression, the fermion contributions $V_f$ can automatically produce a small Higgs VEV,
\bea
  \frac{3 \langle h \rangle}{f} =\frac{\pi}{4} \Rightarrow \xi \approx  (\frac{\pi}{12})^2.
\eea
However, we have seen that in order to reduce the deviation from top Yukawa coupling ($y_t\approx3y'_t$), the Higgs VEV is required to satisfy $3\langle h \rangle/f \ll 1$. To achieve this we need to rely on the contributions from gauge sector to partly cancel $V_f$. Since the Higgs potential is only logarithmically dependent on the partner masses, this cancellation actually doesn't  introduce tuning into the potential. The tuning can be calculated as
\begin{equation}
  \Delta=\left|\frac{\partial\ln\xi}{\partial\ln m_\rho}\right|\approx \frac{1}{\ln\frac{m_\rho^2}{m_W^2}}\Big(\frac{3}{32\xi}-1\Big).
\end{equation}
We can see that the tuning is suppressed by the additional factor $\Big(1/ \log \frac{m_\rho^2}{m_W^2}\Big)$  compared to ordinary Twin Higgs/neutral naturalness models. This is the consequence of the fact that in our model the $Z_2$ symmetry is never explicitly broken. Hence in our model no quadratic dependence on the partner masses is ever reintroduced, leaving the very mild logarithmic sensitivity. Hence for our model the tuning can be significantly suppressed if the gauge and top partners are heavy. In addition, since the Higgs potential is not sensitive to the partner masses, the Higgs mass itself will also be insensitive to them,  thus heavy partners can be achieved with a light Higgs. For example fixing $\xi=0.1/3^2$, $m_h=125$ GeV and $m_t=150$ GeV and using Eqs.(\ref{eq:gamma_fg})-(\ref{eq:Higgs_mass}), we find a roughly estimation of the partners mass scale are around  $M_f\sim3$ TeV, $m_\rho\sim5$ TeV without much tuning $\Delta\simeq 1$. We also show the  numerical values of the tuning in this model for $\xi =0.01$  in Fig.\ref{fig:mh_mf_001} using the measure of tuning from~\cite{Panico:2012uw}.  We can clearly see that no tuning is needed: the  N-suppressed twin Higgs model with TP is fully natural.

For the $N>3$ case, the $s_h^2$ term in $V_f$ is more suppressed, so as we discussed before the vector meson mass can not be as heavy as in the $N=3$ case. In fact it generically turns out to be  below the experimental bound for the realistic case $N\langle h \rangle /f \ll 1$. Hence the $N>3$ case is not interesting: the tuning can not be significantly suppressed and it is also already experimentally excluded from the direct search bounds.

\section{$SO(8)_{V'}$ breaking effects\label{sec:twoloop}}

An essential ingredient of the N-suppression mechanism was the $SO(8)_{V'}$ maximal symmetry, which was a result of the masses of the $\Psi$  being proportional to $V^\prime$. The
$s_{3h}$/$c_{3h}$ structure of the Yukawa couplings was a consequence of this symmetry. However the gauge quantum numbers of the fermions break this symmetry and at two loop order will introduce corrections to the Higgs potential that do not obey the $s_{3h}$ structure.
 Since these fermions are all doublets of $SU(2)_L$ or $SU(2)_L^\prime$ there will be no corrections of this sort from the $SU(2)$ sector. However the various components of $\Psi$ have different
$U(1)_Y \times U(1)_Y^\prime$ hypercharge/hidden hypercharge quantum numbers which at one loop will result in a mass shift that is not proportional to $V^\prime$,
which will break the $SO(8)_{V^\prime}$. However since these fermions mass corrections are only logarithmically sensitive to
the UV scale, if the UV scale is not very high the $SO(8)_{V^\prime}$ breaking effects can be sufficiently small. As we discussed in Sec.~\ref{section:2}, the $V'$ is expected to originate from the VEV of a composite scalar, which should not be generated at scales much higher than the partner masses, hence  the  scale cutting off the log divergences should also not be too high. Since the loop corrections from $U(1)_Y$ and
$U(1)_{Y}^\prime$ are identical, we only focus on the corrections to the fermion masses from ordinary hypercharge. The mass shift that breaks $SO(8)_{V^\prime}$
can be easily calculated from a self-energy diagram:
\bea
\delta M \approx ((\frac{7}{6} )^2  -(\frac{1}{6})^2 )  \frac{3g^{\prime 2}}{(4\pi)^2}M \log \frac{\Lambda^2}{M^2},
\eea
where $7/6$ and $1/6$ are the Hypercharges of the two fermion doublets in the $\bf 4$ of $SO(4)_1 \subset SO(8)$ and $\Lambda$ is the UV cut-off.
Numerically we can estimate for  $\Lambda =100$ TeV and $M =4$ TeV, the ratio  $\delta M/ M \approx 0.02$ hence as expected the
$SO(8)_{V^\prime}$ breaking effects can indeed be very small, of the size of a typical loop factor. Once these maximal symmetry breaking terms are generated at one loop, they will feed into the Higgs potential yielding two loop contributions that will not be $1/N$ suppressed. Since the one-loop contributions are $1/N$ suppressed, these two-loop contributions can give a significant fractional corrections to the entire Higgs potential. However, the most important point is that both the one-loop and the two-loop corrections themselves are very small, the one loop because of the N-suppression, and the two loop just because it is already at two loops.

Let us estimate the leading two-loop correction from the $\delta M$ effects. These will change the effective top Yukawa couplings to take the form
\bea
\frac{M_1 ^t }{\sqrt{2}} \bar{t}_L t_R ( s_{3h} +\frac{\delta M}{M} s_h ) +\frac{M_1 ^t }{\sqrt{2}} \bar{\tilde{t} }_L \tilde{t}_R (-c_{3h} +\frac{\delta M}{M} c_h) +h.c. \nonumber \\
\eea
The resulting leading $SO(8)_{V^\prime}$ breaking contributions in the Higgs potential proportional to $\delta M$ will arise at $\mathcal{O}(y_t^2)$ in
top Yukawa coupling and can be parametrized as
\bea
V_{2-loop} \simeq -c_t^{\prime} \frac{32N_cy_t^{\prime2} f^2 M^2}{(4\pi)^2}\frac{\delta M}{M} s_h^2,
\eea
where $c_t^{\prime}$ is an order one positive number.

 While this correction can be sizeable in comparison to our leading N-suppressed term (for example for $M=4$ TeV and $\Lambda =100$ TeV, $V_{2-loop}$ is around $30 \%$ of the Higgs potential at one-loop), the important point is that overall this is still a very small correction to the Higgs potential which will not significantly increase the tuning needed in the theory. For the above example point we find tuning of order $\Delta \sim 3$.  Increasing to  $M=10$ TeV, $V_{2-loop}$ will become the same order of magnitude as our one-loop term, and result in tuning around $\Delta \sim 10$ for $\xi =0.01$.  The next contribution at two-loop level is at $\mathcal{O}(y_t^2 g^4)$ which is less than $10 \%$ of the one-loop potential and thus can be  safely neglected.
Thus our one-loop N-suppressed Higgs potential  is stable against these two-loop corrections, and no  large tuning is needed as long as the bare mass is not very big and the UV cut-off scale not too high.

\section{Comments on Phenomenology of N-suppression}
\label{sec:pheno}

Finally we would like to highlight the aspects of our $N$-suppression model that could potentially lead to experimentally testable consequences, ultimately leading to a verification of the suppression mechanism, compared to ordinary composite twin Higgs models. 
\begin{itemize}
\item Unbroken twin parity

 The most important novel aspect of $N$-suppression is that twin parity is not broken in gauge and top sector, while in other twin Higgs models it must be broken for successful EWSB. Measuring the Higgs couplings to the top (gauge) sector and the corresponding twin sectors can be used to probe whether or not twin parity is broken.

\item Additional top partners and enahnced global symmetry

The $N$-suppression model presented here requires an additional  multiplet of (twin) top partners, which has an $SO(8)$ global symmetry. These ingrediants are needed for the additional suppression of the Higgs potential. In ordinary twin Higgs models the remaining global symmetry is only  $SO(7)$. Hence the detection of an additional top partner multiplet along with an enhanced global symmetry  could be another good direction for testing our model. 

\item Enhanced Higgs coupling deviations in top sector 

Another novel aspect of the phenomenology of our model is that the coupling of the Higgs to the top sector, $y_t^\prime f \bar{t} t s_{3h}$, is different from the coupling to the gauge sector, $g f^2 W^\mu W_\nu s_h$. On the other hand in ordinary twin Higgs models the pattern of Higgs couplings to the  top and gauge sectors are the same, both proportional to $s_h$. So the Higgs coupling deviation to the top, $\delta y_{tth} \sim N^2 \xi $,  is enhanced by a factor of $N^2$ compared to the gauge sector, $\delta g_{WWh} \sim \xi$ in our $N$-suppression model. Hence  precision measurements of Higgs coupling at the LHC and future colliders could also provide an important for the test of the $N$-suppression mechanism.
\end{itemize}

\section{Conclusions\label{sec:conclusions}}

We have introduced the $N$-suppression mechanism for THM's where the exchange symmetry in the fermion sector acts as an $N$-trigonometric parity. The main consequence is the $1/N^2$ suppression of the quadratic term in the Higgs potential, leading to a natural EWSB model without tuning, while leaving the $Z_2$ symmetry intact. The mechanism for $N=2 n+1$ can be simply implemented by introducing $n$ Dirac fermions with maximal symmetry for each of them, along with some discrete symmetries forbidding unwanted mixing terms. We have presented the $N=3$ benchmark model in detail. Since all dependence on the partner masses is logarthmic, the one-loop Higgs potential is very mildly senitive to them, and we can obtain very small values of $\xi\sim 0.01$ with partner masses $M_f \sim 3$ TeV and $m_\rho \sim 5$ TeV. At two loops we find $N$-suppression violating gauge corrections, however these can be sufficiently small as long as the fermion masses and the cutoff scale are not too large.
We also briefly commented on novel aspects of phenomenology in our model, including the exactly unbroken twin parity, an additional top partner with enhanced global symmetry and enhanced deviations of the Higgs coupling to the top sector.

\section*{Acknowledgements}
 C.C.  thanks the Aspen Center for Physics - supported in part by NSF PHY-1607611 - and the KITP at UC Santa Barbara for its hospitality while working on this project. T.M. thanks the Cornell Particle Theory group for its hospitality while working on this project. C.C. is supported in part by the NSF grant PHY-1719877 as well as the BSF grant 2016153. T.M. is supported in part by project Y6Y2581B11 supported by 2016 National Postdoctoral Program for Innovative Talents in China and also supported by the Israel Science Foundation (Grant No. 751/19), the United States-Israel Binational Science Foundation (BSF) (NSF-BSF program Grant No. 2018683) and the Azrieli foundation. J.S. is supported by the NSFC under grant No.11647601, No.11690022, No.11675243 and No.11761141011 and also supported by the Strategic Priority Research Program of the Chinese Academy of Sciences under grant No.XDB21010200 and No.XDB23000000.

\appendix
\section{The Gauge Sector of the Twin Higgs Model\label{App:gauge}}

\begin{figure}
  \centering
  \includegraphics[width=6cm]{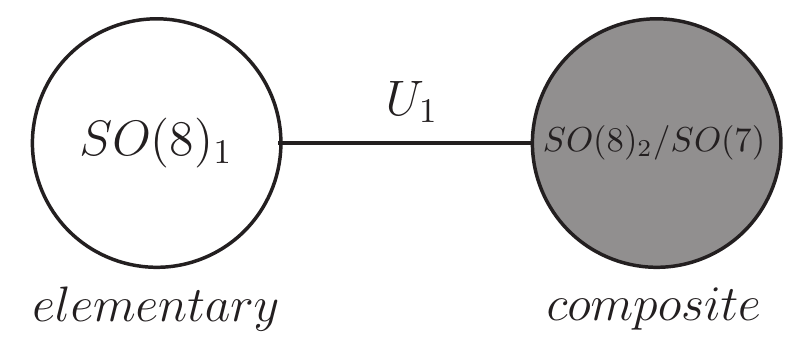}\\
  \caption{Two site model for the $SO(8)/SO(7)$ gauge sector.}\label{fig:gauge_moose}
\end{figure}

In this Appendix we review the structure of the gauge sector of our model. It turns out to be almost identical to the structure of ordinary $SO(8)/SO(7)$ two-site THM's, with the added complication that the $SO(8)_2$ gauge symmetry at second site
is broken by two scalars: one in the symmetric representation of $SO(8)_2$ with VEV $V$ and the other one in the fundamental representation of $SO(8)_2$ with VEV $\mathcal{V}$, which will result in additional uneaten
NGBs. However these additional NGBs can get heavy masses from operators that explicitly
break their shift symmetry (see App.~\ref{app:B}) and can decouple at
low energies where the gauge sector will be identical to those of ordinary two-site $SO(8)/SO(7)$ twin Higgs model as in Fig.\ref{fig:gauge_moose}. Without losing generality,
in the following, we can just use the
ordinary two-site model to calculate the gauge contributions to the Higgs
potential.
   At the first (elementary) site, the SM $SU(2)_L \times U(1)_Y$ and the twin $SU(2)_L^\prime \times U(1)_Y^\prime$ gauge groups are embedded in $SO(4)_1$ and $SO(4)_2$, where the $SO(4)_{1,2}$ subgroups acts on the first (last) four indices of $SO(8)_1$. The $U_1$ link field in the bi-fundamental representation of the global symmetry connects the two sites.  At the second (composite) site we gauge the entire $SO(8)_2$. In order to realize the $SO(8)/SO(7)$ coset space, this gauge symmetry should be broken by a scalar in the fundamental representation of $SO(8)_2$ with VEV $\mathcal{V}$, leading to a nonlinear sigma field $U'$ of the $SO(8)_2/SO(7)$ coset. Since the $SO(8)_2$ gauge symmetry is broken, some NGBs will be eaten by the $SO(8)_2$ gauge bosons which become massive. The uneaten NGBs are contained in $U\equiv U_1U'$ which describes the $SO(8)_1/SO(7)$ coset. The twin gauge symmetry is broken by the global symmetry breaking VEV $\mathcal{V}$ to the twin $U(1)_{em}^\prime$ and the EW gauge symmetry is broken by the pNGB Higgs VEV to the ordinary  $U(1)_{em}$. After EWSB the only uneaten NGB is the physical Higgs boson contained in $U$. We parametrize the NGB fields as
\bea
  U_1=\text{Exp}(\frac{i\pi_1^aT^a}{f}),\quad U'=\text{Exp}(\frac{i\pi_2^{\hat{a}}T^{\hat{a}}}{f}),
\eea
where the $T^a$'s  are the $SO(8)$ generators while the $T^{\hat{a}}$'s are the broken generators in the $SO(8)/SO(7)$ coset, with the normalization   $\text{Tr}[T^aT^b]=\delta^{ab}$. The $Z_2$ invariant gauge interactions of these NGB fields  have the form
\begin{align}\label{eq:gauge_Lagarangian}
  \mathcal{L}_g&=\frac{f^2}{2}\text{Tr}[D_\mu U_1(D^\mu U_1)^\dag]+f^2(D_\mu^\prime \mathcal{H}')^\dag D^{\prime \mu}\mathcal{H}'\nonumber\\
  &-\frac{1}{4}\text{Tr}[ \rho_{\mu\nu}\rho^{\mu\nu}]-\frac{1}{4}W_{\mu\nu}^a W^{\mu\nu,a}-\frac{1}{4}B_{\mu\nu}B^{\mu\nu}\nonumber\\
  &-\frac{1}{4}\tilde{W}_{\mu\nu}^a \tilde{W}^{\mu\nu,a}-\frac{1}{4}\tilde{B}_{\mu\nu}\tilde{B}^{\mu\nu},
\end{align}
where $\mathcal{H}'\equiv U'\mathcal{V}$ is the linearly realized sigma field and the $\rho_\mu=\rho_\mu^aT^a$ are the $SO(8)_2$ gauge bosons. The covariant derivative is $D_\mu U_1=\partial_\mu U_1-i[g (A_\mu^aT_L^a+\tilde{A}_\mu^a\tilde{T}_L^a)+g' (B_\mu T_R^3+\tilde{B}_\mu\tilde{T}_R^3)]U_1+ i g_\rho U_1\rho_\mu^aT^a$ and $D_\mu^\prime \mathcal{H}'=\partial_\mu \mathcal{H}'-ig_\rho \rho_\mu^aT^a \mathcal{H}'$, where $T_L^a$($\tilde{T}_L^a$) and $T_R^3$($\tilde{T}_R^3$) are the  SM and twin generators embedded into the $SO(8)$. Notice that the $Z_2$ symmetry requires the gauge couplings of the SM and twin sectors to be equal. One can easily check that the Lagrangian (\ref{eq:gauge_Lagarangian}) is invariant under the $Z_2$ exchange symmetry defined as
\begin{align}
  &\mathbb{A}_\mu\leftrightarrow P\tilde{\mathbb{A}}_\mu P,\;U_1\rightarrow PU_1,\;U'\rightarrow U'P_0\nonumber\\
  &\Rightarrow A_\mu^a\leftrightarrow \tilde{A}_\mu^a,B_\mu\leftrightarrow \tilde{B}_\mu,s_h\leftrightarrow c_h,
\end{align}
where $\mathbb{A}_\mu\equiv g A_\mu^aT_L^a+g'B_\mu T_R^3$, $\tilde{\mathbb{A}}_\mu\equiv g\tilde{A}_\mu^a\tilde{T}_L^a+g'\tilde{B}_\mu\tilde{T}_R^3$ and
\begin{equation}
  P=\left(
      \begin{array}{cc}
         & \mathds{1}_4 \\
        \mathds{1}_4 &  \\
      \end{array}
    \right),\;P_0=\left(
                    \begin{array}{cccc}
                       &  & \mathds{1}_3 &  \\
                       & -1 &  &  \\
                      \mathds{1}_3 &  &  &  \\
                       &  &  & 1 \\
                    \end{array}
                  \right).
\end{equation}
This $Z_2$ is just an exchange symmetry between the SM EW  and its twin sector which includes the Higgs trigonometric parity (TP), so the Higgs potential induced by the gauge sector must be TP invariant. After integrating out the heavy resonances at tree level, the $SO(8)_1$ invariant effective Lagrangian at leading order in gauge fields can be expressed as (in momentum space)
\begin{align}\label{eq:effe_gauge}
\mathcal{L}_g^\text{eff}=\frac{P_t^{\mu \nu}}{2}\Big[&-p^2\big(W^a_\mu W^a_\nu+B_\mu B_\nu
\big)\nonumber\\
&+ \Pi_0(p^2)\text{Tr}[A_\mu A_\nu]+ \Pi_1(p^2) \mathcal{H}^T A_\mu A_\nu \mathcal{H}\nonumber\\
&+\big(W_\mu^a,B_\mu,s_h\rightarrow \tilde{W}_\mu^a,\tilde{B}_\mu,c_h\big)\Big],
\end{align}
where $\mathcal{H}=U\mathcal{V}$, $P_t^{\mu\nu}=g^{\mu\nu}-p^\mu p^\nu/p^2$ is the projector on transverse field configurations and $A_\mu=g W_\mu^a T_L^a+g^\prime B_\mu T_{R}^3$. The expression of the form factors $\Pi_{0,1}$ are
\bea \label{eq:form_factor}
\Pi_0 = \frac{ p^2 f_\rho^2 }{p^2 -m_\rho^2}, \quad  \Pi_1 = f^2 + \frac{2p^2f_a^2}{p^2 -m_a^2}- \frac{2p^2f_\rho^2}{p^2 -m_\rho^2}.
\eea
The EW and twin EW gauge bosons mass can be extracted from the effective Lagrangian:
\begin{equation}
m_W=\frac{gf}{2}s_h,\:m_Z=\frac{m_W}{\cos\theta_w},\:m_{\tilde{W},\tilde{Z}}=m_{W,Z}(s_h\rightarrow c_h),
\end{equation}
where $\theta_w$ is the Weinberg angle. The TP invariant Higgs potential after integrating out the gauge bosons from Eq.~(\ref{eq:effe_gauge}) is
\begin{align}
V_g &= \frac{3}{2}  \int \frac{d^4p_E}{(2\pi)^4} \Big(2 \text{log}\Big[1+\frac{\Pi_1}{\Pi_0^W} \frac{s_h^2}{4} \Big]  \nonumber \\
&+ \text{log}\Big[ 1 +(\frac{\Pi_1}{ \Pi_0^W} +\frac{\Pi_1}{\Pi_0^B } ) \frac{s_h^2}{4} \Big] \Big) + (s_h \to c_h),
\end{align}
where $\Pi_0^W = -p^2/g^2 + \Pi_0$ and $\Pi_0^B =\Pi_0^W(g \to g')$. Since the Higgs potential is invariant under TP, the $\mathcal{O}(g^2)$ corrections to the potential will be proportional to $s_h^2+c_h^2$ and thus vanish. Thus the leading order Higgs potential is at $\mathcal{O}(g^4)$ proportional to $s_h^4 +c_h^4$ and has the form
\bea
V_g = -\gamma_g (s_h^4 + c_h^4),
\eea
with
\bea\label{eq:gamma_g}
\gamma_g = \frac{9}{64(4\pi)^2} \int^{\infty}_{m_W^2} dp_E^2 p_E^2 \frac{\Pi_1^2 }{(\Pi_0^W)^2},
\eea
where we neglected the hypercharge gauge coupling $g'$.

\section{The discrete symmetries for general $N$  \label{App:discrete}}
\label{app:B}
In this final Appendix we show how to assign the discrete symmetries in the general case to ensure we only get the right mixing terms of Fig.~\ref{fig:Yukawa} needed for the N-suppression.
\begin{align} \label{eq:mixing}
\mathcal{L} &= \sum_{i=1}^n M_i \bar{\Psi}_{i, L} V^\prime \Psi_{i, R} + y_0 f \bar{\Psi}_{q_L}\Sigma'\Psi_{1,R} \nonumber\\
&+\sum_{i=1}^{n-1} y_i f \bar{\Psi}_{i, L} \Sigma' \Psi_{i+1, R}+y_n f \bar{\Psi}_{n, L} \mathcal{H} t_R.
\end{align}

We assume that each elementary fermion $\Psi_{i}$ as well as $q_L$ and $t_R$ is odd under its own $Z_2^i$ symmetry (with $i=0$ corresponding to $q_L$ and $i=n+1$ to $t_R$). This will ensure that none of the elementary fermions can directly mix with each other.

Next we show how they will mix via Yukawa couplings. We assume that the $SO(8)/SO(7)$ coset space can be realized through the two-site  model in Fig.~\ref{fig:gauge_moose}. Each site has a global $SO(8)$ and the $SO(8)_2$ at the second site is fully gauged. The link field $U$ is in the bi-fundamental representation of $SO(8)_1 \times SO(8)_2$. At the second site, there are $n$ scalar fields $\Phi_i$ ($i=1,\ldots, n$) which are in the symmetric representation of $SO(8)_2$ and a scalar $\Phi'$ which is in the fundamental representation of $SO(8)_2$. These scalars break $SO(8)_2$ to $SO(7)$ with VEV $V_i = \text{diag}(1_7, -1 )$ and $\mathcal{V}=(0_7,1)$. We also assume $\Phi_i$ is odd under $Z_2^{i-1} \times Z_2^i$ and $\Phi'$ is odd under $Z_2^n \times Z_2^{n+1}$. The uneaten NGBs then can be described by the linearly realized sigma field $\Sigma_i^\prime =U\Phi_i U^\dagger$ and $\mathcal{H}=U\Phi'$ transforming under $SO(8)_1$ as $\Sigma_i^\prime \to g \Sigma_i^\prime g^\dagger$ and $\mathcal{H} \to g\mathcal{H}$ with $g \in SO(8)_1$. Now we can see that the $\Sigma_i^\prime$ and $\mathcal{H}$ are odd under $Z_2^{i-1} \times Z_2^i$. Since there are $n+1$ sets of uneaten NGBs in the coset space $SO(8)_1/SO(7)$, we can add mass terms for the $n$ sets of NGBs in the form
\bea
\sum_{i=1}^{n-1} f^2_i (\text{Tr}[\Sigma_i^\prime \Sigma_{i+1}^\prime] )^2+f^2_n(\mathcal{H}^\dag\Sigma_n^\prime\mathcal{H})^2.
\eea

So finally find that only one set of the NGBs is massless which results in $\Sigma_i^\prime =\Sigma_{i+1}^\prime=... \equiv \Sigma'$ and this can play the role of the SM Higgs. Since the $\Psi_i$ is odd under $Z_2^i$ and $\Sigma_i^\prime$ is odd under $Z_2^{i-1} \times Z_2^i$, the $Z_2^N$ invariant terms will be

\begin{align}
\mathcal{L} &= \sum_{i=1}^n M_i \bar{\Psi}_{i, L} V^\prime \Psi_{i, R} + y_0 f \bar{\Psi}_{q_L}\Sigma_1^\prime\Psi_{1,R} \nonumber\\
&+\sum_{i=1}^{n-1} y_i f \bar{\Psi}_{i, L} \Sigma_{i+1}^\prime \Psi_{i+1, R}+y_n f \bar{\Psi}_{n, L} \mathcal{H} t_R.
\end{align}

This was we find that the  interactions of the massless NGBs are exactlt as expected in Eq.~\ref{eq:mixing}.

\end{document}